\newif\iflongpaper

\longpapertrue

\documentclass[a4paper]{llncs}

\usepackage{amsmath}
\usepackage{amssymb}
\usepackage{amstext}
\usepackage{amsfonts}
\usepackage{stmaryrd}
\usepackage{subcaption}
\usepackage{graphicx}
\usepackage{paralist}
\usepackage{boxedminipage}
\usepackage{listings}
\usepackage{wrapfig}
\usepackage{color}
\usepackage[american]{babel}
\usepackage{graphicx}
\usepackage[normalem]{ulem}
\usepackage{xspace}
\usepackage{hyperref}

\newcommand\SCION{{\small\textsf{SCION}}\xspace}

\newcommand\A{\textsf{A}\xspace}
\newcommand\B{\textsf{B}\xspace}
\newcommand\C{\textsf{C}\xspace}
\newcommand\D{\textsf{D}\xspace}

\newcommand\F{\textsf{F}\xspace}
\newcommand\G{\textsf{G}\xspace}
\renewcommand\H{\textsf{H}\xspace}
\newcommand\I{\textsf{I}\xspace}

\newcommand\cI{\ensuremath{\mathcal{I}}\xspace}
\newcommand\cJ{\ensuremath{\mathcal{J}}\xspace}

\makeatletter
\let\@copyrightspace\relax
\makeatother

\begin{document}

\title{SCION Five Years Later:\\Revisiting Scalability, Control, and Isolation \\on Next-Generation Networks}

\author{David Barrera, Raphael M. Reischuk, Pawel Szalachowski, Adrian Perrig}
\institute{Network Security Group\\ETH Z\"{u}rich}

\maketitle

\thispagestyle{plain}

\section{Introduction}

The Internet has been successful beyond even the most optimistic
expectations. It permeates and intertwines with almost all aspects of
our society and economy. The success of the Internet has created a
dependency on communication as many of the processes underpinning the
foundations of modern society would grind to a halt should
communication become unavailable. However, much to our dismay, the
current state of safety and availability of the Internet is far from
being commensurate given its importance.

Although we cannot conclusively determine what the impact of a
1-minute, 1-hour, 1-day, or 1-week outage of Internet connectivity on
our society would be, anecdotal evidence indicates that even short
outages have a profound negative impact on governmental, economic, and
societal operations. To make matters worse, the Internet has not been primarily
designed for high availability in the face of malicious actions by
adversaries. Recent patches to improve Internet security and
availability have been constrained by the current Internet
architecture, business models, and legal aspects. Moreover, some of the
fundamental design decisions of the current Internet inherently
complicate secure operation.

To address these issues, we study the design of a next-generation
Internet architecture that provides a fundamental building block: highly available point-to-point
communication. In addition to availability, the architecture should
offer security by design, it should provide incentives for
deployment, and it should consider economic and political issues at
the design stage.

As a solution to address these desired properties, we propose the
inter-domain network architecture \textbf{\SCION}, which is also an
acronym for Scalability, Control, and Isolation on Next-Generation
Networks. In this article, we present (a retrospective
of) its goals and design decisions, its attacker model and
limitations, and 5 years of research conducted since the initial
publication~\cite{ZhHsHaChPeAn2011}.

\section{Goals}

In this section, we present high-level goals an inter-domain
point-to-point communication architecture should satisfy; we
illustrate why these goals are important and how they can be
achieved. Finally, we briefly discuss non-goals, i.e., specific
properties that we intentionally excluded from the design goals.

\subsection{Availability in the Presence of Adversaries}

\index{property!availability}
\index{availability}

Our overarching goal is the design of a point-to-point communication
infrastructure that remains \emph{highly available} even in the
presence of distributed adversaries: as long as an attacker-free path
between endpoints exists, that path should be discovered and used
with guaranteed bandwidth between these endpoints.

Availability in the presence of adversaries is an exceedingly challenging
property to achieve. An \emph{on-path adversary} may drop, delay, or alter
packets that it should forward, or inject packets into the network.  The
architecture hence needs to provide mechanisms to circumvent these malicious
elements. An \emph{off-path adversary} could launch hijack attacks to attract traffic to flow through
network elements under its control. Such traffic attraction can take various
forms; for instance, an adversary could announce a desirable path to a
destination (e.g., by using forged paths or attractive network metrics). Conversely, the
adversary
could render paths not traversing its network less desirable (e.g.,
by inducing congestion).  An adversary controlling a large botnet could also
perform Distributed Denial of Service (DDoS) attacks, congesting selected
network links.
Finally, an adversary could interfere with the discovery of legitimate paths (e.g,
by flooding path discovery with bogus paths).

\subsection{Transparency and Control}

\index{property!path transparency}
\index{property!path control}
\index{path transparency}
\index{path control}

We aim to provide greater transparency and control for (1)
forwarding paths of network packets, and (2) trust roots that are used
for entity validation.

\subsubsection{Transparency and Control over Forwarding Paths.}

When the network offers path transparency, endpoints know (and can verify) the
forwarding path taken by network packets. Applications that transmit sensitive
data can benefit from this property, as packets can be ensured to traverse
certain Internet service providers (ISPs) and avoid others.

Taking transparency of network paths as a first property, we aim to
additionally achieve path control, a stronger, more influential property, with
which receivers can control incoming paths through which they are reachable.
Given a path to a receiver, senders can control end-to-end paths.
This seemingly benign requirement has various repercussions --
beneficial but also fragile if implemented incorrectly. The beneficial
aspects of path control for senders and receivers include:
\begin{compactitem}

  \item \emph{Separation of network control plane and data plane.}
  To enable path control, the control plane (which determines
  networking paths) needs to be separated from the data plane (which
  forwards packets according to the determined paths). The separation
  ensures that forwarding cannot retroactively be influenced by
  control-plane operations, e.g., routing changes. Moreover, the
  separation contributes to enhanced availability as working
  forwarding paths cannot be disrupted by routing changes, but it
  also requires mechanisms to deal with link failures.

  \item \emph{Enabling of multi-path communication.} 
  Path control empowers multi-path communication by letting senders
  select multiple paths to carry packets towards their destinations.
  Multi-path communication is a powerful mechanism to enhance
  availability~\cite{AnBaKaMo2001}.

  \item \emph{Defending against network attacks.}
  If the packet's path is carried in its header (which
  is one way to achieve path control), then the destination can reverse the
  packet path to return its response to the sender, mitigating amplification
  attacks. Path control also enables circumvention of malicious network entities
  or congested network areas, providing a powerful mechanism against DoS and
  DDoS attacks.

\end{compactitem}
\smallskip
The fragile aspects that need to be handled with particular care are
the following.
\begin{compactitem}

  \item \emph{Respecting ISPs' forwarding policies.} If senders have complete
      path control, they may violate ISPs' forwarding policies.  We thus need
      to ensure that ISPs offer a set of policy-compliant paths amongst which
      senders can choose from.

  \item \emph{Preventing malicious path creation.} A malicious sender could
      exploit path control for attacks, for example by forming malicious
      forwarding paths such as loops that consume increased network resources.

  \item \emph{Scalability of path control.} Source routing does not scale to
      inter-domain networks, as a source would need to know the network
      topology to determine paths. To make path control scale, we ensure that
      sources select amongst a relatively small set of paths.  We thus rely on
      source-selected paths instead of full-fledged source routing.

  \item \emph{Permitting traffic engineering.} Fine-grained path control would
      inhibit ISPs from operating and performing traffic engineering. We thus
      seek to provide autonomous system (AS) level path control only at the
      level of ingress/egress interfaces, allowing ISPs to still control paths
      internally.

              \end{compactitem}

\smallskip
\subsubsection{Transparency and Control over Trust Roots.}
Roots of trust are used for the
verification of entities in today's Internet. For example, verification of the
server's public key in a TLS certificate, 
or verification of a Domain Name Service (DNS) response in DNSSEC~\cite{rfc4033}.
Transparency of trust roots provides the property that an end host or user can
know the complete set of trust roots that it needs to rely upon for the
validation of an entity certificate. Such enumeration of trust roots is
complicated today because of intermediate Certification Authority (CA)
certificates that are not explicitly listed but implicitly trusted, e.g., in
the TLS public key infrastructure (PKI). In fact, independent studies have
counted over 300 roots of trust in the TLS
PKI~\cite{SSL-observatory,AbBiMiWoXi2013}, but because of the lack of
transparency there may be additional ones these studies have missed.

Providing control for trust root selection enables \emph{trust
agility}~\cite{moxie-trust-agility}, allowing \emph{users} to easily select or
exclude the roots of trust they want to rely upon.
The challenge then becomes the validation of each certificate, regardless of the choice
of trust roots by users and network entities (e.g., web servers).

\subsection{Efficiency, Scalability, and Extensibility}

\index{property!efficiency}
\index{property!scalability}
\index{efficiency}
\index{scalability}

Despite the lack of availability and transparency, today's Internet
also suffers from a number of efficiency and scalability
deficiencies: for instance, the Border Gateway Protocol (BGP), a global inter-domain routing protocol,
encounters scaling issues in cases of network fluctuations,
where routing protocol convergence can require minutes~\cite{SahKanMoh2009}.
A 2006 earthquake in Taiwan that severed several undersea
communication cables caused Internet outages throughout Asia for several days~\cite{taiwan-earthquake-bbc}.
Moreover, routing tables
have reached the limits of their scalability due to prefix de-aggregation (i.e.,
announcement of more specific prefixes) and multihoming~\cite{Hust2015}.
Unfortunately, extending the memory size of routing tables is
challenging as the underlying Ternary Content-Addressable Memory
(TCAM) hardware is expensive and power-hungry, consuming on the order
of a third of the total power consumption of a router. Extending the
routing table memory would thus drastically increase cost and power
consumption of routers.

Security and high availability come at a cost, usually resulting
in lower efficiency and potentially diminished scalability. High
performance and scalability, however, are required for viability in the current
economic environment. We therefore explicitly seek \emph{high efficiency} as a
goal in the common case such that packet forwarding latency and
throughput are at least as fast as current IP forwarding. Moreover,
we seek \emph{improved scalability} compared to the current Internet,
in particular with respect to BGP and the size of routing
tables.

An approach to achieve efficiency and scalability is to
avoid router state wherever possible. We observe that modern
block ciphers such as AES can be computed faster than
performing memory lookups. For example, on current PC platforms,
computing AES requires on the order of 50 cycles while fetching a byte
from main memory requires around 200 cycles~\cite{intelaesniwhitepaper}. 
Moreover, modern block ciphers can
be implemented in hardware with tens of thousands of gates, which is
sufficiently small to replicate it profusely, which in turn enables high
parallelism -- the high complexity of a high-speed memory system prevents such
replication at the same scale.
We thus aim to
place state into packet headers and protect the state cryptographically,
enabling higher packet processing speeds 
and simpler router architectures compared to today's IP routers.
Besides higher efficiency, avoiding state on routers also prevents
state-exhaustion attacks~\cite{SVMFHK2011} and state inconsistencies.

Our goal of efficiency and scalability is in line with the \emph{end-to-end
  principle}, which states that a function should be implemented at the network layer
in which it can most effectively operate~\cite{SalReeCla84}. Since the end host
has the most information about its internal state, functions such as bit error
recovery, duplicate suppression, or delivery acknowledgments are best handled
by the end host. Compared to the current Internet, \SCION applies the end-to-end
principle one layer lower in the protocol stack. Currently, most
transport-layer functionality is handled by the end host according to this
principle. However, in \SCION, end hosts also assist with network-layer
functionality such as path selection. End host path selection is communicated to
the network by packet-carried forwarding information, which in turn removes the
need for inter-domain routing tables at border routers. Consequently, end
host path selection results in a simpler forwarding plane and thus more
efficient routers.

To future-proof \SCION, we design the core architecture and code base to be
\emph{extensible}, such that additional functionality can be easily built and
deployed. \SCION clients and routers should (without overhead or expensive
protocol negotiations) discover the minimum common feature set supported by all
intermediate nodes.

\subsection{Support for Global but Heterogeneous Trust}

\index{property!global trust}
\index{property!heterogeneous trust}
\index{global trust}
\index{heterogeneous trust}

Given the diverse nature of constituents in today's Internet
with diverse legal jurisdictions and interests, an important
challenge is how to scale authentication of entities (e.g., AS ownership for
routing, name servers for DNS, or domains for TLS) to a global environment. The
roots of trust of currently prevalent PKI models (monopoly and oligarchy) do not
scale to a global environment because mutually distrusting entities cannot agree
on a single trust root (monopoly model), and because the security of a plethora
of roots of trust is only as strong as its weakest link (oligarchy model). We
thus seek a trust architecture that supports \emph{meaningful trust roots in a global
environment} with mutually distrusting entities.

\subsection{Deployability}

\index{property!deployability}
\index{deployability}

Incentives for deployment are important to
overcome the resistance for upgrading today's Internet. A multitude of features is
necessary to offer the initial impulse: high availability even under
control-plane and data-plane attacks (e.g., built-in DDoS defenses),
path transparency and control, trust root transparency and control, high efficiency,
robustness to configuration errors, fast recovery from failures, high forwarding efficiency, multi-path forwarding, etc.

If early adopters cannot obtain sufficient benefits from migrating to
a new network architecture, even initial deployment is unlikely to be
successful. Ideally, even the first deploying ISP can gain a
competitive advantage, and start selling services to its initial
customers.

Migration to the new architecture should require minimal added complexity to the
existing infrastructure. Deployment should be possible by re-utilizing the
internal switching infrastructure of an ISP, and only require installation or
upgrade of a few border routers. Moreover, configuration of the new architecture
should be similar to the existing architecture, such as in the configuration of
BGP policies, minimizing the amount of additional personnel training. 

Economic and business incentives are also of critical importance. ISPs should be
able to define new business models and sell new services. Users should derive a
business advantage from the new architecture, for example, obtain properties
similar to a leased line at a fraction of the cost. Migration cost should be
minimal, requiring only the deployment of low-cost routers. Finally, a new
architecture should not disrupt current Internet business models, but maintain
the current Internet topology and business relationships (e.g., support
peering).

\subsection{Non-Goals}

We deliberately exclude certain properties and goals
that could be added as additional functionality later on. For example, we do not consider
multicast or efficient content dissemination as part of the basic
communication infrastructure, as we recognize the significant
complexity these features would add. Also, these features can be effectively
added through an overlay leveraging a next-generation Internet architecture's
basic communication infrastructure~\cite{FLTGKMNSS2013}.

\medskip
We additionally consider several other problems out of scope for a network
architecture.
A major category of current security problems are software
vulnerabilities. While software vulnerabilities of end hosts are
clearly out of scope, software vulnerabilities of network components such as
routers can
affect network operation. It is thus important to address these
network vulnerabilities through robustness to malicious components
and attempts to reduce them through a simple network architecture.
Malicious Internet content (e.g., spam or phishing emails, malicious
web pages) should not be directly addressed by the communication infrastructure.
The architecture, however, should offer mechanisms that assist in
defending against theses threats.

\section{SCION Architecture Overview}

We now provide a high-level overview of the \SCION architecture. A more detailed
description is in papers available on our web site \url{http://www.scion-architecture.net}.

\begin{figure*}[tp]
  \centering
  \includegraphics[width=.9\textwidth]{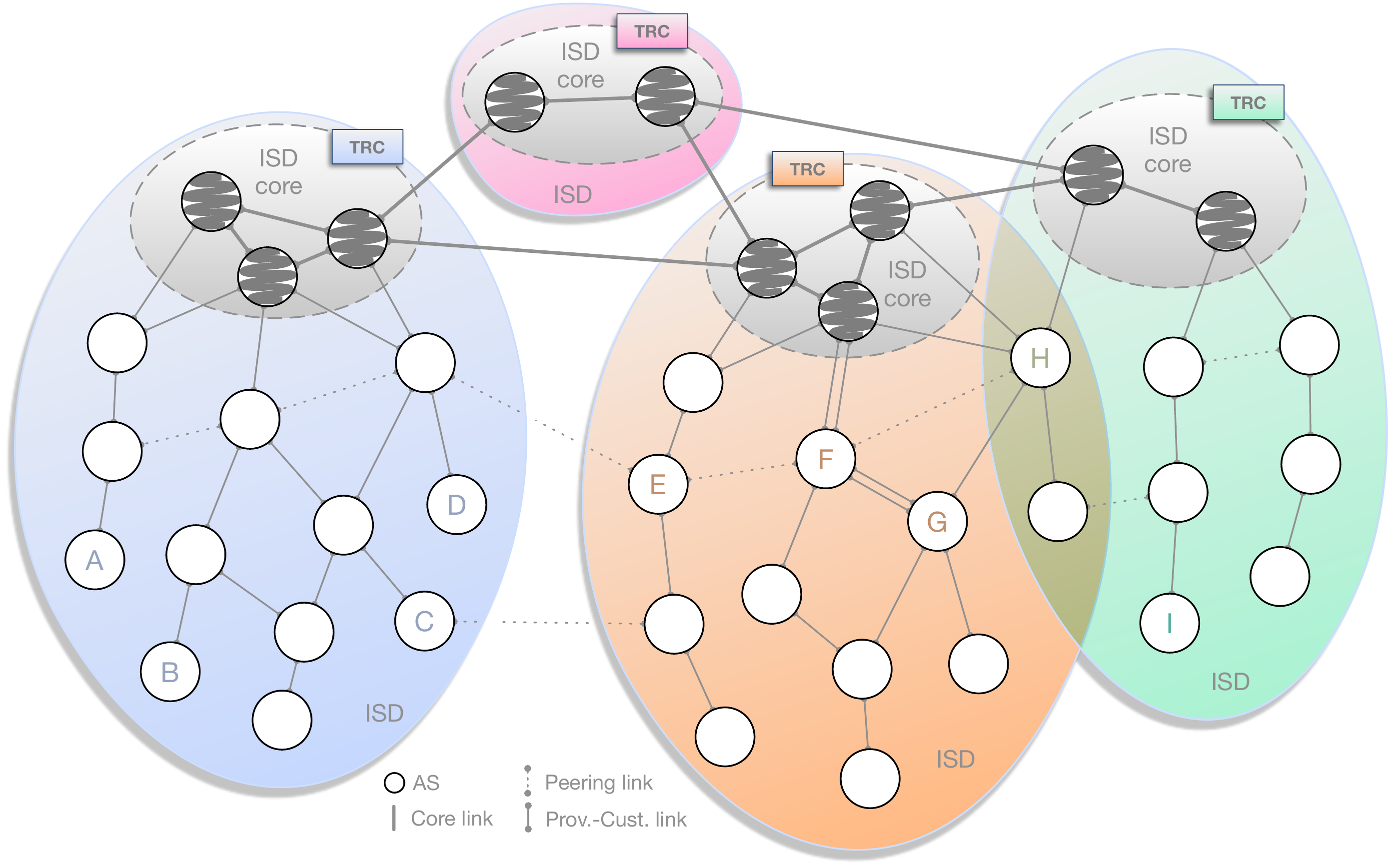}
  \caption{Four\,\SCION ISDs with ISD Cores and ASes. The ISD Core
  ASes are connected via Core paths. Non-Core ASes are connected via
  customer-to-provider or peering links. AS \H is contained in two
  ISDs.}
  \label{fig:ISDs}
\end{figure*}

\subsection{SCION High-Level Overview}

A fundamental building block to achieve the properties of high availability,
transparency, scalability, and support for heterogeneous trust is
\textbf{ISolation Domains (ISDs)}. An ISD constitutes a logical grouping of a
set of \textbf{Autonomous System (ASes)}, as depicted in \autoref{fig:ISDs}.
An ISD is administered by one or multiple ASes, which form the
\textbf{ISD Core}. We refer to these ASes as ISD Core ASes or simply Core
ASes. An ISD contains one or multiple regular ASes.
The ISD is governed by a policy, called
\textbf{Trust Root Configuration (TRC)}, which is negotiated by the
ISD Core. The TRC defines the roots of trust that are used to
validate bindings between named entities and their public keys (key
certificates) or their addresses (DNS). As part of the TRC, every ISD
has an associated human-understandable name space, which is globally
unique. The only global coordination that is required in \SCION is
hence the ISD name and number.

ASes join an ISD by purchasing service from another AS in the ISD; joining an
ISD thus constitutes an acceptance of the ISD's TRC file. Otherwise, they would
select an ISP which is part of an ISD they desire to belong to. Typically, 3--10
current Tier-1 ISPs would constitute the ISD's Core ASes, and their associated
customers would participate in the ISD. We envision that ISDs will span areas
with uniform legal environments that provide enforceable contracts. If two ISPs
have a contract dispute they cannot resolve by themselves, such a legal
environment can provide an external authority to resolve the dispute.  All ASes
within an ISD also agree on the TRC, i.e., the entities that operate the trust
roots and set the ISD policies. ISDs will thus likely be formed along national
boundaries or federations of nations, as entities within a legal jurisdiction
can enforce contracts and agree on a TRC.
ISDs are hierarchical, as \SCION supports sub-ISDs. ISDs can also overlap in the
sense that an AS may be part of several ISDs. Although an ISD does provide
isolation from other networks, the central purpose of an ISD is to provide
transparency and to support heterogeneous trust environments.  Although ISDs may
seem to lead to ``Balkanization'' and prevent an open Internet, they
counter-intuitively provide openness and transparency, as we hope to elucidate in
this article.

\begin{figure*}[tp]
  \centering
  \includegraphics[width=.98\textwidth]{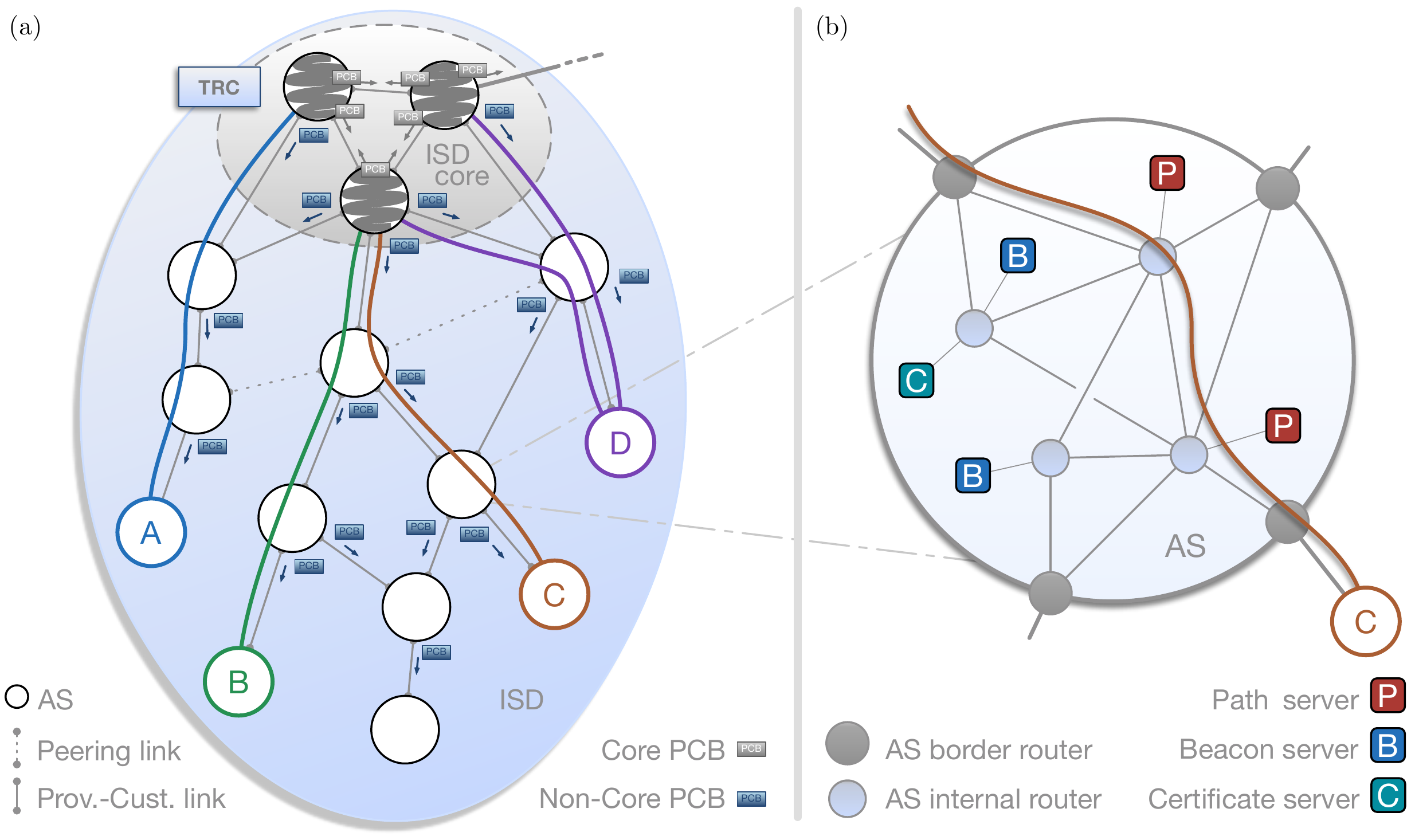}
  {\phantomsubcaption\label{fig:paths}}
  {\phantomsubcaption\label{fig:AS}}
  \caption{
    (\subref{fig:paths}) \SCION ISD with path construction beacons
    (PCBs) that are propagated from the ISD Core down to customer
    ASes, and path segments for ASes \A, \B, \C, and \D to the ISD
    Core.
    (\subref{fig:AS}) Magnified view of an AS with its routers and
    servers. The path from AS \C to the ISD Core traverses two
    internal routers.
  }
\end{figure*}

\SCION uses two levels of routing, intra-ISD and inter-ISD. Both levels utilize
\textbf{Path Construction Beacons (PCBs)} to discover and establish routing
paths (see \autoref{fig:paths}). An ISD Core AS announces a PCB and
disseminates it as a policy-constrained multi-path flood either \emph{within} an
ISD (to discover intra-ISD paths) or \emph{amongst} ISD Core ASes (to discover
inter-ISD paths). PCBs accumulate cryptographically protected AS-level path
information as they traverse the network. These cryptographically protected
contents (that we call opaque fields as described below) within received PCBs are chained together by
sources to create a data transmission path segment that traverses a sequence of
ASes. Packets thus contain AS-level path information
avoiding the need for border routers to maintain inter-domain routing tables.
We refer to this concept as \textbf{Packet-Carried Forwarding State (PCFS)}.

Through the inter-domain PCB transmission process, Core ASes learn
paths to every other Core AS. Through the intra-domain PCB
dissemination, ASes learn path segments on how to reach ISD Core ASes, which
enable an AS to communicate with the ISD Core. \autoref{fig:paths}
shows some path segments from the ASes \A, \B, \C, and \D to the ISD core.

We emphasize that PCFS in \SCION is \textit{different from} source
routing, as a source node does not search a network topology graph to
select its path. Instead, with the approach of
\textit{source-selected paths}, a source node combines at most three
path segments (up-segment, core-segment, and down-segment). Since an
arbitrary source up-segment combined with an arbitrary destination
down-segment (along with an appropriate core-segment if necessary)
results in a valid end-to-end path, a source node does not need to
search any topology to find a path, thus, the approach is
fundamentally different from source routing.

\subsection{Control Plane: Beaconing for Route Discovery}

We now discuss the control plane components and mechanisms in more
detail. The control plane is responsible for discovering paths and
making those paths available to end hosts. \autoref{fig:AS} shows the
main components that perform these operations in \SCION:
\emph{beacon servers} discover path information; \emph{path
servers} disseminate such path information; and \emph{certificate servers} assist with validating received information. Border routers provide the connectivity between ASes.

\begin{figure}[h]
  \centering
  \includegraphics[height=120mm]{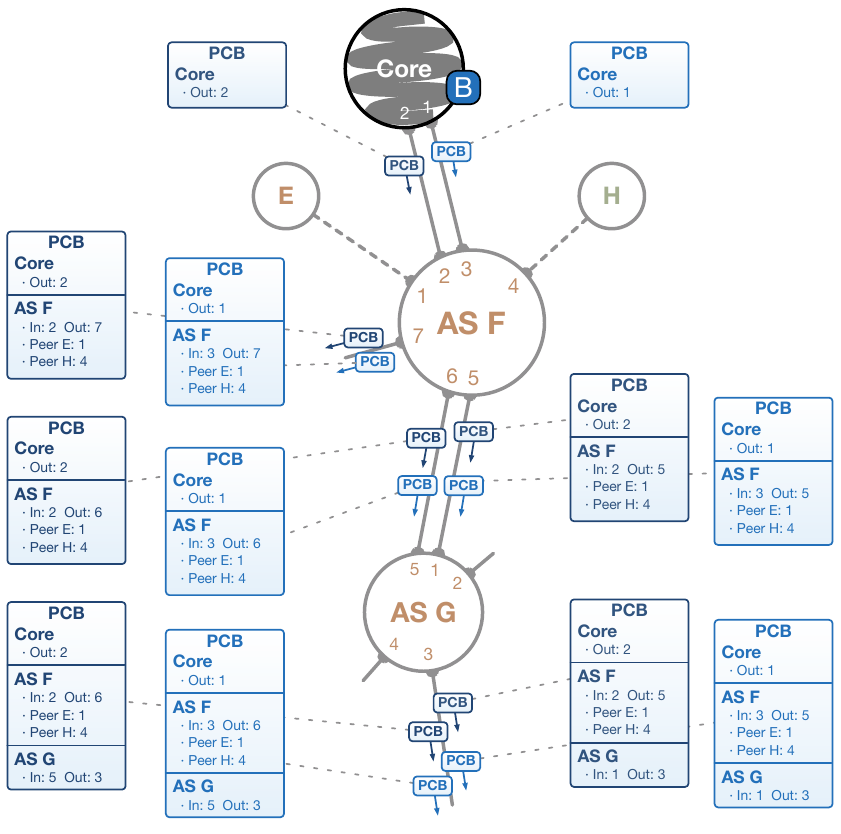}
  \caption{Intra-ISD PCB propagation from the ISD Core down to customer ASes.
  For the sake of illustration, the interfaces of each AS are
  numbered with consecutive integer values. In practice, each AS can choose
  any encoding for its interfaces. In particular, only the
  AS itself needs to understand its encoding.}     \label{fig:PCBs}
\end{figure}

\textbf{Beacon servers} are responsible for the dissemination of PCBs
(see \autoref{fig:paths}). Beacon servers in a Core AS generate
intra-ISD PCBs that are sent to all non-Core ASes of the ISD.
Non-Core AS beacon servers receive such PCBs and re-send them to
their customer ASes, which results in policy-compliant AS-level paths.
\autoref{fig:PCBs} shows PCBs that are propagated from the ISD Core
down to customer ASes. At every AS, information about the interfaces
of the AS is added to the PCB.

The beacon servers run a fault-tolerant protocol to ensure state
consistency across all local servers. Periodically, a master beacon server
generates a set of PCBs that it forwards to its customer ASes. In the case of
inter-ISD communication, the beaconing process is similar to BGP's route advertising process,
although the process is periodic and PCBs are flooded multi-path over
policy-compliant paths to discover multiple paths between any pair of ASes.
\SCION{}'s beacon servers can be configured to express all BGP policies, as well
as additional properties (e.g., control of upstream ASes) that BGP cannot
express.

\textbf{Path servers} store mappings from AS identifiers to sets of such
announced path segments, and are organized as a hierarchical caching system similar to
today's DNS. ASes, through the master beacon servers, select the set of path segments
through which they want to be reached, and upload them to a path server in the
ISD Core.

\textbf{Certificate servers} keep cached copies of TRC files retrieved from the
ISD Core, keep cached copies of other ASes certificates, and manage keys and certificates for
securing intra-AS communication. Certificate servers are queried by beacon
servers when validating the authenticity of PCBs (i.e., when a beacon server does not have
a corresponding certificate).

\smallskip
An AS typically receives several PCBs representing several diverse
path segments on how to reach various ISD Core ASes.
\autoref{fig:paths} shows two path segments for AS \D. We call a
path segment that leads \emph{towards} an ISD Core an \textbf{up-segment}, and a path segment that
leads \emph{from} the ISD Core to an AS a \textbf{down-segment} -- although path segments
are typically bi-directional and thus support packet forwarding in both
directions.  More precisely, up-segments and down-segments are invertible: by
flipping the order, an up-segment is converted to a down-segment and vice versa. Path
servers learn up-segments by extracting them from PCBs they obtain from the local
beacon servers. To reach its ISD Core, a host performs a path lookup at its
local path server, fetching up-segments to the ISD Core. To reach a remote
destination, a host additionally queries its path server for the
down-segment of the destination AS. In case the local path server has
no cached entry for the down-segment, it will query the destination
AS's Core path server.

How do the Core path servers know the down-segments of the destination?
The beacon servers in an AS select the down-segments through
which the AS desires to be reached, and register these path segments at the
Core path servers. When links fail, segments expire, or better segments
become available, the beacon servers keep updating the down-segments
registered for their AS.

End-to-end communication is enabled by a combination of up to three
path segments that form a \SCION end-to-end path. More precisely, the
source host in ISD $\cI$ sends a path resolution request to its local
path server, who forwards the request to a core path server. In case
the requested path's destination is \emph{within} the ISD~\cI, the
core path server responds by immediately sending up to $k$
down-segments to the local path server. In case the requested path's
destination is \emph{outside} the ISD~\cI, then the core path server
first requests the corresponding down-segments from the core path
server in destination ISD~\cJ before responding to the local path
server. In both cases, the local path server returns up to $k$ up-
and down-segments to the requesting source, and if needed, a
core-segment connecting the core of $\cI$ with the core of $\cJ$. Depending on the returned
segments, \SCION paths can be created as follows:

\begin{itemize}

\item Case 1 (\textbf{immediate path segment combination},
e.g., path \B $\to$ \D in \autoref{fig:paths}):
the last AS on the up-segment (ending at a Core AS) is the same AS as
the first AS on the down-segment (starting at a Core AS). In this case,
the simple combination of up- and down-segment creates a valid
end-to-end path.

\item Case 2 (\textbf{AS shortcut},
e.g., path \B $\to$ \C in \autoref{fig:paths}):
the up-segment and down-segment intersect at a non-Core AS. This is the
case of a shortcut where up-segment and down-segment meet before
entering the ISD Core. In this case, a shorter path is possible by
removing the extraneous part of the path. The special case when the
source's up-segment contains the destination AS is treated in the same
way, i.e., the intersection of both segments is omitted from the path.

\item Case 3 (\textbf{peering shortcut},
e.g., path \A $\to$ \B in \autoref{fig:paths}):
a peering link exists between the two segments, so a shortcut via the
peering link is possible. As in Case~2, the extraneous path segment
is cut off. The peering link could be traversing to a different ISD.

\item Case 4 (\textbf{core-segment combination}, e.g., path \A $\to$
\D in \autoref{fig:paths}, or \A $\to$ \I in \autoref{fig:ISDs}): the
last AS on the up-segment is different from the first AS on
down-segment. This case requires an additional core-segment to
connect the up- and down-segment. In case the communication remains
within the same ISD (\A $\to$ \D), a local ISD core-segment is
needed; otherwise (\A $\to$ \I), an inter-ISD core-segment is
required.

\end{itemize}

Once an end-to-end path is chosen, this path is encoded in the \SCION
packet header, which makes inter-domain routing
tables unnecessary for border routers: both the egress and the
ingress interface of each AS on the path are encoded as PCFS in the
packet header. The destination can respond
to the source by inverting the end-to-end path from the packet
header, or it can perform its own path lookup and construction as the
source did.

\SCION's beaconing process has several important aspects. The periodicity is on
the order of 10 seconds -- in the current system a fresh set of beacons is sent
over each inter-AS link to the neighboring ASes every 15 seconds. This beacon
propagation process is thus asynchronous, i.e., PCBs are sent based on a local
timer and are not propagated immediately upon arrival. The paths for propagation
are selected based on a path quality metric with the goal of identifying
consistent, diverse, efficient, and policy-compliant paths.  \emph{Consistency}
refers to the requirement that there exists at least one property along which
the path is uniform, such as an AS capability (e.g., anonymous forwarding) or link
property (e.g., low latency).
\emph{Diversity} refers to the set of paths that are announced over time being
as path-disjoint as possible to provide high quality multi-path options.
\emph{Efficiency} refers to the length, bandwidth, latency, utilization, and
availability of a path, where more efficient paths are naturally preferred.
\emph{Policy compliance} refers to the requirement that the path adheres to the
AS's routing policy.  Based on past PCBs that were sent, a beacon server scores
the current set of candidate path segments and sends the $k$ best segments as the next PCB.
To provide some concreteness to this description, we currently use $k=5$, and
send PCBs every 15 seconds to each neighbor over each provider-to-customer link.  \SCION
intra-ISD beaconing can scale to networks of arbitrary size, because each
inter-AS link carries the same number of PCBs regardless of the number PCBs
received by the AS.

Unlike in the current Internet, link failures are not automatically resolved by
the network, but require more active handling by end hosts.  Since \SCION
forwarding paths are static, they break when one of the links fails. Link
failures are handled by a three-pronged approach that typically masks link
failures without any outage to the application and rapidly re-establishes fresh
working paths. More precisely, 
(1) PCB dissemination occurs every few seconds, constantly establishing new working paths
in case existing paths become unavailable. 
(2) ICMP-like control messages rapidly erase path segments with broken links from path
servers and beacon servers -- thereby triggering the dissemination of additional
PCBs. Beacon servers then immediately send 
additional working paths after
learning of a path failure.  
(3) Most importantly, \SCION end hosts use multi-path communication by default,
thus masking link failures to an application with another working path.  As multi-path communication is very successful in achieving high
availability (even in environments with very limited path
choice~\cite{AnBaKaMo2001}), \SCION beacon servers actively attempt to create
disjoint paths, \SCION path servers make an effort to select and announce
disjoint paths, and end hosts make an effort to compose path segments to achieve
maximum resilience to path failure.  Consequently, most path failures in \SCION
are imperceptible to the application, unlike the numerous short outages plaguing
the current Internet~\cite{KusKanKat2007,KSCCVFMAK2012}.

Paths are represented at AS-level granularity, which by itself is
insufficient for diversity; ASes often have several diverse
connection points, and thus a disjoint path is possible despite the
AS sequence being identical. For this reason, \SCION encodes AS
ingress and egress interfaces as part of the path, exposing a finer
level of path diversity. \autoref{fig:PCBs} demonstrates this
feature: AS \F receives two different beacons via two different links
from the Core. Moreover \F uses two different links to send two
different beacons to AS \G, each containing the respective egress
interfaces. AS \G extends the two beacons and forwards both of them
over a single link to its customer.

An important optimization point is that \SCION also supports peering
links between ASes. Consistent with AS policies in the current Internet, PCBs do
not traverse peering links. However, peering links are announced along with a
regular path in a PCB. \autoref{fig:PCBs} shows how AS$_1$ includes its two
peering links in the PCB. If the same peering link is announced on two paths, then
the peering link can be used for the end-to-end path. \SCION also supports
peering links that cross ISD boundaries, which highlights the importance of
\SCION{}'s path transparency property; a source knows the exact set of ASes and
ISDs traversed during the delivery of a packet.

Communication within an AS is handled by existing intra-domain communication
protocols, such as IP, Multi-Protocol Label Switching (MPLS), or Software-Defined
Networking (SDN) -- border routers encapsulate the \SCION packet inside an IP,
MPLS, or SDN frame.  \autoref{fig:AS} shows one possible intra-domain path from
AS \C to the ISD core.

Inter-ISD beaconing operates similarly to intra-ISD beaconing, except that
inter-ISD PCBs only traverse ISD Core ASes. The same path selection metrics
apply, where an AS attempts to forward the set of most desirable paths to its
neighbors. A difference, however, is that an AS forwards $k$ beacons
\textit{per} source AS, with $k=3$. The periodicity is also reduced, we forward
PCBs once a minute or upon path changes. Similar to BGP, this process is
inherently not scalable, however, as the number of ISDs and the corresponding
number of Core ASes is small, this approach is viable.

\paragraph{Security Aspects}
For protection against malicious ASes and to provide a secure control plane, 
\SCION is equipped
with an arsenal of cryptographic mechanisms. We describe an overview in this
paper; the details are in companion papers~\cite{ZhHsHaChPeAn2011,MRSKP2015}. 

The root of
trust of an ISD is composed of root key certificates of trusted ISD Core ASes and
Certification Authorities (CAs). The ISD's TRC specifies which root key
certificates are trusted and how many different signatures are required for each
operation. For example, an AS certificate may require signatures from 2 different
entities, an update to a new TRC may require signatures from 4 different
entities.

The \SCION control plane includes the \emph{\SCION Control Message Protocol} (SCMP).
One challenge in the design of SCMP was how to enable efficient authentication of SCMP messages,
as the na{\"i}ve approach of adding a digital signature to SCMP messages
could create a processing bottleneck at routers when many SCMP messages would be created in
response to a link failure.  We thus make use of an efficient symmetric key
derivation mechanism called Dynamically Recreatable Key
(DRKey)~\cite{KBJLHP2014}. In DRKey, each AS uses a local secret key known to
\SCION border routers to derive on-the-fly a per-AS secret key using an
efficient Pseudo-Random Function (PRF).  Hardware implementations of modern
block ciphers enable faster computation than a memory lookup from DRAM, and
therefore
such dynamic key derivation can even result in a speedup over fetching the key
from memory. For verification of SCMP messages, the destination AS can fetch the
derived key through an additional request message from the originating AS, which is protected by a
relatively slow asymmetric operation. However, local caching ensures that this
key only needs to be fetched infrequently, about once per hour. As a
consequence, \SCION provides fully secured control messages with minimal
overhead.

Similar to BGPSEC~\cite{LepTur2013}, each AS signs the PCB it forwards. This
signature enables PCB validation by all entities. To ensure path correctness,
the forwarding information within each Packet-Carried Forwarding State (PCFS) also needs to be cryptographically
protected, however, signature verification would hamper efficient forwarding.
Thus, each AS uses a secret symmetric key that is shared among beacon
servers and border routers and is used to efficiently compute a Message
Authentication Code (MAC) over the forwarding information. The per-AS
information includes the ingress and egress interfaces, an expiration
time, and the MAC computed over these fields, which is all encoded within an
8-byte field that we refer to as \textbf{Opaque Field (OF)}. \index{Opaque
Field (OF)}\index{OF|see{Opaque Field}} We use the term opaque because the structure of the
field is largely at the discretion of each AS and requires no
coordination with any other AS -- as long as the AS itself can extract how to
forward the packet on to the next AS.

The specified ingress and egress interfaces uniquely identify the links to the
previous and following ASes. If a router is connected via the same outgoing
interface to 3 different neighboring ASes, 3 different egress interface
identifiers would be assigned. The OF's expiration time can be set on the
granularity of seconds or hours, depending on the type of path. For the
discussion of this overview, we only consider the common case where paths are
long-lived and OFs have an expiration time on the order of 12 hours.

In terms of cryptographic mechanisms, we built in algorithm agility,
such that cryptographic methods can be easily updated and exchanged.
The MAC validation of OFs is per-AS, so an AS can independently
(without interaction with any other entity) update its keys or
cryptographic mechanisms. We support multiple signatures by an AS,
thus, an AS can readily deploy a new signature algorithm and start
adding those signatures as well. The path consistency component of
the beacon selection metric (as explained above) will automatically
start creating paths where each AS supports the new algorithm,
enforcing consistency of the signature type. Validating inter-domain
PCBs is accomplished by requiring connected ISDs to cross-sign their
respective TRC files -- consequently, any sequence of ISDs has a
verifiable sequence of signatures, the details of which are described
in our paper on the SAINT system~\cite{MRSKP2015}.

\subsection{Data Plane: Packet Carried Forwarding State (PCFS)}

While the control plane is responsible for providing end-to-end paths, the data
plane ensures packet forwarding using the provided paths. A \SCION packet
minimally contains a path; source and destination addresses are optional in
case the packet's context is unambiguous without addresses. Consequently,
\SCION border routers forward packets to the next AS based on the AS-level path
in the packet header (which is augmented with ingress and egress interface
identifiers for each AS), without inspecting the destination address and
also without consulting a routing table. Only the border router at the
destination AS needs to inspect the destination address or packet purpose to
forward it to the appropriate local host(s).

An interesting aspect of this forwarding is enabled by the split of locator
(the path towards the destination AS) and identifier (the destination
address)~\cite{rfc6830}: because in-network forwarding does not consider the
local identifier, any source or destination address format is possible. Thus, a
domain can select an arbitrary addressing format for its hosts, e.g., 4-byte
IPv4, 6-byte medium access control, 16-byte IPv6, or 20-byte accountable IP
(AIP~\cite{AIP}) address. A nice consequence is that an IPv4 host could directly communicate with an IPv6 destination.\footnote{Communication between hosts with different network stacks requires support from the host operating system. Today's hosts typically assume compatible stacks on the endpoints.}

Routers can efficiently forward packets in the \SCION architecture. In
particular, the absence of inter-domain routing tables and the absence of
complex maximum prefix matching performed by current routers enables
construction of faster and more energy-efficient routers. During forwarding, a
border router would first verify that the packet entered through the correct
ingress interface. If the packet has not yet reached the destination AS, the
egress interface defines the next hop. For illustration purposes, let us assume
that an AS uses IPv4 switching to internally forward traffic, that the AS has
fewer than 255 border interfaces (ingress- or egress-interfaces), and that each
border interface is directly connected to the neighboring AS. The AS can select
internal addresses such that all border interfaces follow a common addressing
scheme, e.g., the IPv4 address is 10.1.1.X for $0<X<255$. The value X here
uniquely identifies the preceding or following AS, and thus can serve as the
ingress or egress interface identifier in the OF. Consequently, a border router
can simply extract the egress interface identifier X from the OF, encapsulate
the \SCION packet into an IPv4 packet with the destination address of 10.1.1.X,
and let the intra-domain routing and forwarding handle packet delivery to the
egress interface.

\subsection{Entity Validation Infrastructure}

All entity authentication in \SCION is based on traditional certificates, which
bind identifiers to public keys and carry digital signatures that are
verified by roots of trust, i.e., public keys that are axiomatically
trusted.\footnote{The reason we did not make use of self-certifying public
  keys~\cite{HIP,AIP} for long-term identities is because of their
  inherent inability for revocation and the complexities involved with key
  updates. For short-term identities, however, we do appreciate their features.}
The challenge is how to achieve trust agility to enable flexible selection of
roots of trust, resilience to private key compromises, and efficient key
revocation. We explore these issues in detail in our SAINT system~\cite{MRSKP2015},
and provide a high-level overview in this article.

A central question is how to structure the trust roots. Today's Internet follows
two trust models: monopoly and oligarchy. In the \emph{monopoly} model, a single
root of trust is used for authentication. The DNSSEC PKI~\cite{rfc4033} or the
Resource Public Key Infrastructure (RPKI)~\cite{rpki-url} used in BGPSEC are
examples of the monopoly model as they both essentially rely on a single public
key that serves as a root of trust to verify all subsequent entities. The
monopoly model suffers from two main drawbacks: all parties must agree on a
single root of trust, and the single root of trust represents a single point of
failure because its misuse enables forging a certificate for an
arbitrary entity. The \emph{oligarchy}
model does not fare much better -- instead of a single root of trust there are
several roots of trust, all of which are equally and completely trusted.
Instead of one single point of failure in the monopoly model, the oligarchy
model thus exposes several points of failure.
The prime example is the TLS PKI, featuring on the order of 1500
trusted entities including about 300 roots of
trust~\cite{SSL-observatory,AbBiMiWoXi2013}. 
Compromise of a single trusted entity enables forging a server certificate which
allows man-in-the-middle attacks, as we have recently witnessed in several
instances involving Comodo, DigiNotar, and Turktrust.

\SCION's ISDs provide solutions to these issues by allowing each ISD to define its own set of
roots of trust, along with the policy governing their use. Such scoping of trust
roots within an ISD greatly improves security, as compromise of a private key
associated with a trust root cannot be used to forge a certificate outside the
ISD. 

An ISD's trust roots and policy are encoded in the \textbf{Trust Root Configuration
(TRC)} file. The TRC has a version number, a list of public keys that serve as
roots of trust for various purposes, and policies governing how many signatures
are required for different certificates, how many signatures are needed to
update the TRC, etc. The TRC serves as a way to bootstrap all entity
authentications. We now discuss two properties offered by the TRC:
\emph{trust agility} and \emph{efficient revocation of trust roots}.
\textbf{Trust agility} offers the selection of the sets of roots of trust to
initiate validation of certificates. A user can thus select an ISD that
she believes maintains an uncompromised set of trust roots. A challenge
with trust agility is to maintain global verifiability of all entities,
regardless of the user's selection. \SCION offers this property by requiring
all ISDs with a link among them to cross-sign each other's TRC files -- as long
as a network path exists, a validation path thus exists along that network path.
\textbf{Efficient revocation of trust roots} is the second important property.
In today's Internet, trust roots are revoked manually, or through OS or browser
updates, often requiring a week or longer until a large fraction of the
Internet population has observed such revocations. In \SCION, PCBs carry the
version number of the current TRC, and the updated TRC is required to validate
that PCB. An AS that realizes that it needs a newer TRC can contact the AS from
whom it has received the PCB. Following the distribution of PCBs, an entire ISD
updates the TRC within tens of seconds.

\SCION also introduces new mechanisms for the validation of network entities
such as: ASes, path server and DNS responses, or web servers. We separate
authentication into two different types based on their respective emphasis on
either availability or security~\cite{MRSKP2015}. One type is \textbf{routing
authentication} to authenticate PCBs, which has availability as the main
requirement since control plane messages must be authenticated to provide
communication paths. To achieve a higher level of security, additional servers
would need to be contacted to provide resilience to private key compromises,
but this in turn would hamper availability. In some cases, the requirement to
communicate with additional servers introduces a circular dependency, because
routing is required to communicate with these servers, but contacting these
servers is needed to verify the routing message. Therefore, we ensure that all
information required to authenticate routing messages flows in the
same direction as the routing messages themselves, avoiding circular
dependencies. In the case of \SCION{}, AS certificates binding ASes to public
keys flow from providers to customers, in the same direction as PCBs.

The second type is \textbf{service authentication}, which serves the purpose of
authenticating services such as web servers or DNS replies. Since the control
plane is operational when hosts communicate with servers, additional entities
can contribute to ensure higher security to verify server authenticity, such as
integrity log servers and validators in the Attack-Resilient PKI (ARPKI)
system~\cite{BCKPSS2014}. ARPKI is a highly secure PKI system based on log servers
that keep a public log of all certificates to monitor CAs' operations, and CAs
and validators that monitor operation of log servers. By requiring multiple
signatures on certificates, and by adding signatures on all operations we
obtain the property that at least 3 malicious trusted entities within the same
ISD are needed to perform a man-in-the-middle attack on a single domain.
To further increase security, we designed PoliCert, a system to enable domains
to specify their detailed security policy~\cite{SMP2014}. By storing the
domain policies in an ARPKI log, policy consistency and integrity is ensured.
In concert, ARPKI and PoliCert achieve a high level of security for domains'
certificates -- all PKI attacks we have witnessed in the past decade would have
been avoided in this framework.
As a last line of defense, we propose efficient gossip protocols for
verifying the consistency of log servers~\cite{laurent_gossip}. Clients
randomly exchange short information about the logs to guarantee that any
misbehavior will eventually be detected. 

\subsection{Incremental Deployment and Incentives}

Support for incremental deployability plays a key role in the successful
adoption of any network architecture. To this end, we have designed \SCION to
be deployable (by both ISPs and end users) without requiring substantial
changes to the existing infrastructure.

\subsubsection{Incremental Deployment.}
At a minimum, ISPs need only deploy a 
border router capable of encapsulating and decapsulating \SCION traffic as it 
leaves or enters their network. \SCION ASes must also deploy 
certificate, beacon, and path servers. These servers can run on commodity 
hardware and can optionally be replicated for 
increased availability, e.g., \autoref{fig:AS} contains two path
servers and two beacon servers. The current version of the \SCION
codebase uses IP as an underlying protocol, which allows existing
intra-domain networking infrastructure and configuration.

In terms of creation and deployment of ISDs, we envision these to grow
organically, with one ISD initially defined for each area with uniform legal
environments. Tier-1 ISPs within those ISDs would
become Core ASes. \SCION facilitates the evolution of ISD and AS structure
through efficient updates to the TRC file. 

Deploying \SCION to end user sites (e.g., homes or businesses) is designed to
require little effort as well, initially requiring no changes to existing software,
networking stacks, nor the replacement of end user network devices. 
For initial deployment, we achieve customer-friendly deployment through the
design of a device (which we call DENA, or Device for ENhancing Availability)
that can be installed at customer locations. DENA is a bump-in-the-wire
middlebox that sits between the customer's WAN connection and their ISP. The
device transparently monitors network flows and identifies remote flow
endpoints where other DENAs are present. If a remote DENA is discovered for a
given network flow, a \SCION path is established as a fallback channel in case
network communication through standard BGP paths is lost or becomes unreliable.
DENA is able to switch between BGP and \SCION paths transparently (providing
pseudo-multi-path functionality to network applications), improving network
availability for all devices behind the middlebox, and without requiring
changes to networking stacks.

\subsubsection{Deployment Incentives.}
An important issue to consider is what incentives exist for various parties for
deploying \SCION. For end users, the benefits of using \SCION are plentiful,
ranging from higher throughput and lower latency communication (which
translates to better quality phone calls, and higher resolution video streams),
to fewer Internet outages.  Users can also benefit from the \SCION extensions,
enabling, for example, low-latency anonymous communication. 

By deploying \SCION, ISPs can provide high-availability service to their
customers, and simultaneously increase resilience to DDoS attacks for both
themselves and their customers. These service offerings can enable new revenue
streams for ISPs who have deployed \SCION. There is a relatively low cost to
transitioning existing BGP business models and policies to \SCION, as these
policies can be expressed and even extended. In \SCION such policies are less prone to
attacks, but also to configuration errors since any error is constrained to a
local domain.

Similarly to ISPs, businesses (possibly running their own ISDs) can benefit
from highly available communication at lower cost. Deploying \SCION provides better
connectivity for customers, and higher resilience to DDoS. As additional
benefits, path control reduces the possibility of industrial or government
espionage, while transparency further provides deterrence for such practices.
Finally, control over the businesses' PKI can prevent man in the middle
attacks. 

Governments have shown interest in deploying \SCION for a variety of reasons.
The ability to avoid a global root of trust, and to select their own roots,
allows governments to cooperate and trust parties whom they deem fit. The open
nature of the \SCION codebase allows it to be deployed freely onto any supported
(and possibly verifiable) hardware device. This would help governments in cases where
particular hardware vendors cannot be trusted.

As of 2015, we have deployed a global \SCION testbed which we are actively
using to vet \SCION's functionality and security. As of July 2015, the testbed includes
deployment nodes in 5 continents with 3 ISDs and 20 ASes. We are continually
adding nodes at universities and corporate sites. Details and
requirements for sponsoring a \SCION node can be found on our website.

\subsection{Extensions}

\SCION{}'s extensible architecture enables new systems that can take advantage
of the novel properties and mechanisms provided. As compared to the current
Internet, most of the benefits can be afforded through the use of PCFS, path
transparency, and control. We briefly describe noteworthy protocols and systems
that have been built as extensions to \SCION.

\textbf{Path validation} -- \SCION, through its use of PCFS, paves way for the
Origin and Path Trace (OPT) mechanism. OPT enables the sender,
receiver, and routers to cryptographically verify the path that the packet
traversed~\cite{KBJLHP2014}. By leveraging the DRKey mechanism, routers can
efficiently derive their key, verify the path, and update the path validation
fields.

\textbf{Anonymity and privacy} -- PCFS also
provides advantages for privacy.  With PCFS and path transparency, the source
is able to select paths that appear more trustworthy (e.g., those that do not
traverse certain ASes).  In addition, the packet header can be further
obfuscated such that ASes on the path cannot learn identifying details about
the source or the destination, unless they are immediately connected to one of
them. Proposals such as LAP~\cite{HKPYNGM2012} and HORNET~\cite{CABDP2015}
leverage \SCION's path selection infrastructure to offer high bandwidth and low
latency anonymous communication. 

\textbf{DDoS defense} -- the hierarchical organization of ASes into a
manageable number of ISDs enables neighbor-based contracts between pairs of
ISDs, which in conjunction with path segments inside the ISDs allows for
establishing efficient bandwidth guarantees between any two end hosts. Such
bandwidth guarantees are leveraged by SIBRA~\cite{BRSPZHKU2015} to prevent DDoS
attacks at the architectural level: independent of the number of distributed
bots, end hosts obtain protection against Internet-wide link-flooding attacks,
one of the major threats in today's Internet.

\section{Case Studies}

\SCION improves many aspects of the current Internet. This section highlights some of the
use cases that demonstrate unique properties offered by the new architecture.

\textbf{Constraining traffic flows through trusted ASes} --
Path control and transparency are important properties for sensitive traffic,
where a sender wants control and assurance over which ASes will be traversed, 
due to legal, secrecy, or safety considerations. For instance, banking or 
medical data, which is typically bound to strict data privacy regulations, can 
be constrained in \SCION to traverse only selected authorized ASes. Furthermore, 
the OPT mechanism enables a sender and receiver to verify the exact path taken 
on a per-packet basis, with negligible overhead~\cite{KBJLHP2014}.

\textbf{Highly available communication for critical infrastructures} --
Critical infrastructures such as financial networks and industrial control 
systems used for power distribution require a 
high degree of
availability. Internet outages have been known to wreak havoc on 
day-to-day operations, for example preventing ATM withdrawals or payment 
terminal operations~\cite{malaysialeak-bgpmon}. \SCION{}'s control-plane 
isolation through ISDs, its stable data plane, and its multi-path operation all 
contribute to dramatically higher availability.

\textbf{High-speed web browsing} -- 
Current congestion control hinders high-speed communication because the sender
and receiver require time to determine their sending rate and to constantly
perform congestion control.  Consequently, the sending rate is usually below
the maximum possible rate. In \SCION, through the SIBRA~\cite{BRSPZHKU2015}
extension, the sender performs a resource reservation with its initial packet,
and the receiver will likely obtain a reservation with a high sending rate that
it can immediately start to use on the reverse path. On such a reservation, no
congestion control is needed; consequently, the web server can immediately
start sending the web page at a high rate to the browser.

\textbf{Mobility support} --
With the proliferation of mobile devices, supporting reliable communication can
be challenging since these devices frequently connect and disconnect from 
(sometimes several) networks. \SCION supports high availability through 
multi-path communication and provides a header extension to inform the other
party of new down segments. In \SCION,
a mobile device that obtains a new address or connection as it connects to
a new network can send new down segments to the other party.
Failing paths are discarded and new paths are dynamically discovered
transparently
to users and applications. One challenging case, however, is when both sender and 
receiver simultaneously move to a 
new network and all the previously established paths fail at the same time. In 
this infrequent case, a name resolution server needs to be contacted to fetch
fresh down segments for the other party~\cite{STUVWY2014}.

\section{Attacks and Defenses}

\SCION dramatically improves network security as compared to the current
Internet, which we illustrate based on three important classes of attacks and their defenses.

\textbf{Prefix hijacking} -- Numerous Internet outages are due to the malicious or
erroneous announcement of IP address space, which is also known as prefix
hijacking. Perhaps the most famous case is the hijack of YouTube by Pakistan for
internal censoring, resulting in a global outage of YouTube. In fact, hijacks
that impact only a small portion of the Internet happen on a daily basis. \SCION
prevents such hijacking through several mechanisms. With ISDs, misconfigurations and attacks in 
one ISD do not automatically affect others; digitally signed route 
announcements mean unauthorized injection of routes is not possible; and
digitally signed path distribution allows verification of paths by the sender.

\textbf{Forged TLS certificates} -- Compromised roots of trust have been used to
create rogue TLS certificates. A famous case is where the government of Iran
used forged certificates for Google and Yahoo services to perform
man-in-the-middle attacks on its citizens; Iran is suspected to have mounted
the attack on the DigiNotar CA, who signed these certificates. The ISDs and the
ARPKI system used in \SCION prevent such attacks, as a CA's authority is scoped to
the ISDs where the CA is active in, and using ARPKI at least 3 trusted entities
all need to be compromised to perform a successful man-in-the-middle attack.
Moreover, the \SCION root of trust update mechanism enables revocation of roots
of trust within tens of seconds, enabling quick recovery from
compromise.

\textbf{DDoS attacks} -- Large-scale DDoS
attacks have been widely used to prevent access to domains. For example, a
large-scale attack against Estonia made several of their critical
infrastructures inaccessible during one week in April
2007~\cite{estoniajournal}. \SCION would have minimized the impact of these
attacks. ISDs allow external traffic to be de-prioritized, thus enabling
internal communication in case the attack originates outside the ISD. Critical
infrastructures can keep some network paths to a destination secret, thus
preventing an adversary from even sending traffic to that destination because
the cryptographic OFs are necessary to use a path.  The SIBRA extension offers
powerful mechanisms for DDoS defense, as it guarantees minimal traffic rates
between any pair of ASes, which cannot be lowered even by a large-scale
botnet~\cite{BRSPZHKU2015}, even when using new types of DDoS attacks such as
Crossfire and Coremelt~\cite{StuPer2009,KanLeeGli2013}.

\section{Deployment Caveats}

The allocation and structure of ISDs presents a challenge for the deployment of
\SCION. It remains unclear, for example, which ASes within an ISDs will or
should become Core ASes. We envision that among a group of ASes who deploy a
top-level ISD, the AS or ASes that can form peering agreements 
with core ASes in other ISDs should become core ASes in their own ISD.
However, \SCION itself does not require or impose strict rules regarding the
allocation of ISDs; ISDs can overlap, which means an AS can belong to several
ISDs (cf.\ AS \H in \autoref{fig:ISDs}). Sub-ISDs are possible as well, offering
the flexibility to start an ISD without needing to peer with core ASes of other
ISDs and enabling finer-grained control over routing isolation and
authentication.  In this context, the important properties \SCION offers are
path control and transparency: as long as communicating hosts can select and
inspect the path of their packets, the question of ISD partition is of secondary
nature.

Another challenge that could arise is that each AS will attempt to be its own
ISD or will want to be part of the ISD Core.
While too many top-level ISDs will pose a problem for \SCION scalability,
we observe that economically sound decisions will lead to larger ISDs due to
economies of scale -- because the startup costs and operation of an ISD Core AS
are more expensive than a non-Core AS, the operation of a large ISD will amortize
the cost over more non-Core ASes. Moreover, ASes preferentially associate with
larger ISDs, which can offer better connectivity to other ISDs as well as to
other ASes within the ISD. On the other hand, ISD growth is limited to only as
large as entities can agree on the ISD's TRC (i.e., roots of trust). Finally,
ASes desiring to be part of the ISD Core are assessed in the same way current
ASes assess peering: an AS is permitted into the Core if the current Core ASes
deem it to be large enough to fulfill Core AS duties (e.g., participating in
beacon and path server replication as illustrated in \autoref{fig:AS}).

As expected in architectures with PCFS, packet headers are necessarily larger.
Larger headers place a bound on goodput, since payload space is traded off for
header space. The current \SCION codebase implements the OF as an 8-byte field.
Since every AS on an end-to-end path has to be represented through a corresponding OF,
the overhead increases linearly with the number of ASes on the path.
However, given that the average AS path in today's Internet is four hops long (and
decreasing)~\cite{caida_twelve,avg_as}, the overhead introduced by SCION should
not exceed 40-50 bytes on average.  The performance penalty of transmitting more
packets appears reasonable since per-packet forwarding performance is
faster than routing-table-based architectures. While the default
header size has not shown a performance disadvantage in our testing, many of the
proposed \SCION extensions add length to the header.

Certain extensions (e.g., SIBRA, HORNET) have been designed for a use case
assuming pervasive deployment (i.e., deployment at a majority of \SCION ASes).
While the benefits of these extensions are clear, we must consider the
efficiency implications of certain ASes not deploying the extensions. For
example, data payload space may be lost due to additional signatures or key
material for path validation on nodes that do not validate paths, leading to
inefficiency in data transfer. We have designed our extensions to be compatible
with non-deploying nodes, but future work should consider improvements such as
opportunistic enabling of these extensions.

Due to path dissemination and registration dynamics, \SCION beacon and path servers can
incur a high overhead under specific circumstances. For example, if a given
link's state fluctuates frequently between available and unavailable (due to
error, hardware fault, or an adversary), the beacon server would need to
consistently update the set of paths that include that link, and serve new
paths excluding that link. We expect that this case will be rare, but also
easily detectable.  Additionally, higher quality (uptime, availability) links
will have higher probability of selection, minimizing the impact of rapid path
fluctuations.

We have shown that the basic building blocks of \SCION are relatively
straightforward to understand and have many beneficial properties and
applications. However, as more extensions and alternative PKIs
are added to the architecture, the operational complexity of
the architecture increases correspondingly. We believe that this additional
complexity is
worth the security, efficiency, and availability guarantees provided
by the extensions. It is ultimately up to the networking and research community
to decide which of these extensions will be deemed worthwhile for widescale deployment.

\section{Related Work}
\label{sec:related}

Several efforts on redesigning the Internet have been made over the past two
decades to satisfy the new requirements of emerging Internet-based
applications. Such requirements include naming, routing, mobility, network
efficiency, availability, and evolvability of the Internet. We discuss several
projects in this space based on a loosely temporal order clustered by topics.

The idea of clustering the network into domains has been attempted since the
early days of the Internet. The Nimrod routing architecture~\cite{rfc1992}, to
our knowledge, is the first published description of these concepts.  Nimrod
describes a hierarchy of clusters of hosts, routers, or networks. A secure
version of Nimrod was later proposed \cite{SirKen97}.  FARA~\cite{Clark03FARA}
proposes a general notion of an entity to include clusters of computers that
can be reached as a communication endpoint.

The NewArch project~\cite{Clark03FARA} describes comprehensive requirements for
a new Internet, such as separation of identity from location, late binding
using association, identity authenticity, and evolvability. However, it mostly
emphasizes a new direction for end-point entities while the packet delivery in
the current IP network is left intact.  NewArch uses the New Internet Routing
Architecture (NIRA)~\cite{YanClaBer2007} for inter-domain routing, which aims
to introduce competition among ISPs in the core by providing route control to
the end users who can choose domain-level paths.

NDN~\cite{Jacobson09ccn,ndn-url} decouples location from identity and uses
identity for locating the corresponding content. NDN relies heavily on
in-network caching of data and is useful for accessing popular static content.
However, NDN's scalability would suffer in the face of an increasing number of
new, ephemeral content (e.g., voice or video calls), and require even more
complex and energy-consuming routers than IP routers.  The Publish-Subscribe
Internet Routing Paradigm (PSIRP) project supports information-centric
networking based on a publish-subscribe approach~\cite{PSIRP}. They propose an
elegant approach to reduce the state on routers by having packets carry Bloom
filters to encode the next hops of a multicast packet~\cite{JZEAN2009}.  The
CCNx project provides a specific implementation of content-centric networking,
developing detailed specifications and prototype systems~\cite{ccnx}.

Mobility-first~\cite{RaNaVe2012} is an architecture that quickly maps billions
of identities to their locations, yet does not propose a fundamental change in
the underlying forwarding architecture in terms of security and availability.
Nebula~\cite{Nebula} addresses security problems in the current Internet.
Nebula takes a so-called default-off approach to reach a specific service,
where a sender can send packets only if an approved path to a service is
available. The network architecture helps the service to verify whether the
packet followed the approved path (i.e., supporting path verification).
However, Nebula achieves this property at a high cost. All routers on the path
need to perform computationally-expensive path verification for every single
packet and need to keep per-flow state, limiting its usage to highly
specialized services. 

Serval~\cite{NSGKAKRF2012} proposes name-based service discovery and routing,
and introduces a new service-access layer that enables late binding of a
service to its location. Late binding provides flexibility in migrating and
distributing services, yet it attempts to optimize networking for special
application-services (especially in data-center network) built on top of the
current Internet. 

XIA~\cite{Han12xia} proposes an evolvable network architecture that can easily
adapt to the evolution of networks by supporting various principal types (where
the principal includes but is not limited to service, content, host, domains,
and path). Due to its flexibility, yet lack of specific data-plane mechanism,
XIA uses \SCION for secure and available data forwarding.  

All the aforementioned new Internet architectures attempt to solve issues
facing applications built on top of the current Internet, yet do not address
the very fundamental architectural problems that hamper available and private
data communication in the presence of malicious parties.  The Framework for
Internet Innovation (FII)~\cite{KSBFGGGMPRRAK2011} also proposes a new
architecture to enable evolution, diversity, and continuous innovation, such
that the Internet can be composed of a heterogeneous conglomerate of
architectures.  The ChoiceNet~\cite{WGCRBN2014} architecture proposes an
``economy plane'' to enable network providers to offer new network-based
services to customers, providing an network environment for improving
innovation and competition.

Several architecture proposals suggest the approach of better path control for
senders and receivers, for example Segment Routing~\cite{segmentrouting},
Pathlets~\cite{GoGaShSt2009}, NIRA~\cite{YanClaBer2007},
i3~\cite{StAdZhShSu2004}, or SNAPP~\cite{PaPeAn2008}.

Forward~\cite{forward-url} and SysSec~\cite{syssec-url} are proposing to build
secure and trusted  Information and Communication Technology (ICT) systems by
engaging academia and industry.  Forward is an initiative by the European
Commission to promote the collaboration and partnership between industry and
academia in their common goal of protecting ICT infrastructures.  The Forward
project categorizes security threats to various ICT systems including
individual devices, social networks, critical infrastructures (such as smart
electric grids), and the Internet infrastructure, and it aims at coordinating
multiple research efforts to build secure and trusted ICT systems and
infrastructures.  SysSec aims to consolidate the systems security research
community in Europe, promoting cybersecurity education, engaging a think-tank
in discovering the threats and vulnerabilities of the current and future
Internet, creating an active research road map in the area, and developing a
joint working plan to conduct state-of-the-art collaborative research.  Since
Forward and SysSec currently focus on identifying and handling
threats, we believe our proposed tasks to be a good addition
to the projects by providing an architecture that would significantly reduce
the attack surface. RINA~\cite{rina-url} is a recursive inter-network
architecture that provides unified APIs across all protocol layers. In RINA,
all layers have the same functions with different scope and range, where a
layer is a distributed application that performs and manages inter-process
communication. We would make an effort to design our prototype to fit into this
paradigm so that our architecture can support seamless integration with other
higher-layer security protocols/mechanisms.

Many researchers are currently studying Software-Defined Networking (SDN), for
example in the OpenFlow~\cite{McKeown08Openflow,openflow-url} project. These
efforts mainly consider intra-domain communication, which \SCION can leverage
to communicate within a domain.

Several future Internet efforts provide testbeds for running and testing a new
architecture, such as GENI~\cite{geni-url}, Fi-ware~\cite{fi-ware-url} and
FIRE~\cite{fire-url}.

We have developed \SCION with a focus on security and high availability for
point-to-point communication, which is a unique perspective and can contribute
to other future Internet efforts. For instance, even content-centric networking
needs a routing mechanism to reach the data source. \SCION can offer the
routing protocol to support that functionality. Once a server is found in a
service-based infrastructure or a nearby content cache is found in a
content-centric architecture, point-to-point communication between the end host
and the server will offer the highest communication efficiency, as pure
forwarding is faster than server-based or content-based lookups. Similarly,
\SCION can provide the point-to-point communication fabric in a
mobility-centric architecture. Consequently, \SCION offers mechanisms that
complement many previously proposed future Internet architectures.

\section{Conclusions}

We have presented \SCION, a future Internet architecture that provides security,
availability, transparency, and scalability. We have demonstrated that \SCION
offers numerous advantages over competing architectures, but can also work
jointly with other proposals as an underlying building block for highly reliable point-to-point
communication.

Despite its research maturity after 5 years of work, \SCION is still in its
infancy in terms of deployment. While requiring relatively small changes by
ISPs and domains, broadening adoption is currently \SCION{}'s greatest
challenge. We expect that the potential benefits for various stakeholders will
provide strong incentives to drive adoption, leading to islands of \SCION
deployment. In the long term, connections and mergers among islands will enable
ever-increasing numbers of native \SCION end-to-end connections.

Working on \SCION has offered us the opportunity to think about Internet
architectures from a clean-slate perspective. 
The absence of limiting constraints (imposed by the current Internet
environment) has been particularly rewarding, as the deep exploration of a
problem space enabled us to design a system with properties that were
previously thought to be impossible. 
We anticipate that the insight into the possible applications of a secure,
dynamic, and highly-available network will help engage the network community to
leverage \SCION for their applications, and contribute to the project.

\section*{Acknowledgements}

We thank the original \SCION authors as well as current and past members of the
ETH Z\"{u}rich Network Security group and of the CMU Security group for their
contributions to the project. We also thank Yih-Chun Hu and Brian Trammell for interesting
discussions, feedback, and suggestions that improved this article.

\bibliographystyle{abbrv}
\bibliography{bib,erc-bib}

\end{document}